\documentclass[11pt]{article}%
\usepackage{amssymb,amsmath,xcolor,graphicx,xspace,colortbl, rotating}
\usepackage{amsfonts}
\usepackage{amsthm}
\usepackage{caption}
\usepackage[mag=1000,portrait]{geometry}
\usepackage{indentfirst}
\usepackage[onehalfspacing]{setspace}
\usepackage{subcaption}
\usepackage{booktabs}
\usepackage{amsmath}
\usepackage{amssymb}
\usepackage{graphicx}%
\usepackage{multirow}
\usepackage{multicol}
\usepackage{amsmath}
\DeclareMathOperator*{\arginf}{arginf}
\usepackage[justification=centering]{caption}
\providecommand{\U}[1]{\protect\rule{.1in}{.1in}}
\DeclareGraphicsExtensions{.pdf,.eps,.ps,.png,.jpg,.jpeg}
\providecommand{\U}[1]{\protect\rule{.1in}{.1in}}

\DeclareMathOperator*{\argmin}{arg\,min}

\begin{document}

\author{$%
	\begin{array}
		[c]{ccccc}%
		\text{Byunghoon Kang}\thanks{b.kang1@lancaster.ac.uk, Department of Ecnomics, Lancaster University}& \hspace{0.2in} & \text{Seojeong Lee}\thanks{s.jay.lee@snu.ac.kr (corresponding author), Department of Economics, Seoul National University}~\thanks{ Lee acknowledges that this work was supported by the New Faculty Startup Fund from Seoul National University.}& \hspace{0.2in} & \text{Juha Song}\thanks{yuyuhee2@snu.ac.kr, Department of Economics, Seoul National University} \\
		\text{Lancaster University} & \hspace{0.2in} & \text{Seoul National University}& \hspace{0.2in} & \text{Seoul National University}\\
	\end{array}
	\medskip$}
\date{\today}

\title{Convergence Rates of GMM Estimators with Nonsmooth Moments under Misspecification}
\maketitle	

\begin{abstract}
The asymptotic behavior of GMM estimators depends critically on whether the underlying moment condition model is correctly specified. Hong and Li (2023, \textit{Econometric Theory}) showed that GMM estimators with nonsmooth (non-directionally differentiable) moment functions are at best $n^{1/3}$-consistent under misspecification. Through simulations, we verify the slower convergence rate of GMM estimators in such cases. For the two-step GMM estimator with an estimated weight matrix, our results align with theory. However, for the one-step GMM estimator with the identity weight matrix, the convergence rate remains $\sqrt{n}$, even under severe misspecification.
\end{abstract}


\section{Introduction}
The generalized method of moments (GMM) is a unifying estimation framework that efficiently combines information about the average behavior of economic variables to estimate parameters of interest defined by the underlying moment condition model. In this note, we focus on over-identified, nonsmooth (non-differentiable), and potentially misspecified moment functions, investigating the convergence rates of both one-step and two-step GMM estimators through extensive simulations.

Under standard assumptions, the convergence rate of GMM estimators is $\sqrt{n}$, regardless of whether the moment function is differentiable (Hansen, 1982) or not (Newey and McFadden, 1994). However, when the moment condition is misspecified, the convergence rate can slow down. Specifically, while GMM estimators remain $\sqrt{n}$-consistent and asymptotically normal when the moment function is smooth (Hall and Inoue, 2003) or directionally differentiable (Kang and Lee, 2024), they become $n^{1/3}$-consistent when the moment function is non-directionally differentiable (Hong and Li, 2023).

Although not specific to GMM, other studies have also demonstrated that the convergence rate of an estimator can change under misspecification. Koo and Seo (2015) investigated this phenomenon in the context of structural break models. When the model is correctly specified or exhibits weak misspecification, such as an incorrect number of breaks, the estimator for the break location converges rapidly to the true breakpoint at a rate of $n^{-1}$. However, under strong misspecification, such as when the true regression function is neither linear nor time-invariant, the oracle property no longer holds, and the convergence rate of the breakpoint estimator drops significantly, reaching at most $n^{-1/3}$.

Similarly, Hidalgo, Lee, and Seo (2019) examined threshold models and found that if the model is continuous (e.g., a kink model) but the true restriction is not imposed during estimation, the convergence rate of the breakpoint estimator is reduced to $n^{-1/3}$.

Although Hong and Li (2023) theoretically establish the slower convergence rate for nonsmooth moments under misspecification, they do not provide explicit simulation results for this\footnote{They only provide simulation results comparing the finite sample performance of their proposed rate-adaptive bootstrap with the standard bootstrap.}. To address this gap, we conduct an extensive simulation study to complement the theoretical results. We employ two simulation designs based on those in Hong and Li (2023): (i) a simple location model and (ii) an IV-quantile regression model.

For the two-step efficient GMM estimator or the one-step GMM estimator with an estimated weight matrix, our simulation results align with the theoretical predictions, showing that the variance of the GMM estimator decreases at a rate of $n^{-2/3}$.\footnote{Let $\widehat{\theta}$ denote the GMM estimator and $V \equiv \text{var}(n^{1/3}(\widehat{\theta} - \theta_{0})) > 0$. Then, $\text{var}(\widehat{\theta}) = V/n^{2/3}$.} However, for the one-step GMM estimator with the identity weight matrix, we observe that the convergence rate remains $\sqrt{n}$, even under severe misspecification. This is unexpected, as Theorem 1 of Hong and Li (2023) establishes the cubic-root convergence rate for GMM estimators with a fixed weight matrix.

\section{Model and Estimator}
The moment condition is given by 
\begin{equation*}
E[g(X_{i},\theta_{0})]=0
\label{mc}
\end{equation*}
for a unique $\theta_{0}$, where $g(X_{i},\theta)$ is a known function of the random variables $X_{i}$ and the parameter of interest $\theta$. The moment condition is just-identified if $\text{dim}(\theta)=\text{dim}(g(x,\theta))$ and over-identified if $\text{dim}(\theta)<\text{dim}(g(x,\theta))$. An over-identified moment condition model is misspecified if
\begin{equation*}
	E[g(X_{i},\theta)]\neq0,~\forall\theta.
	\label{mis}
\end{equation*}
It is worth emphasizing that this type of moment misspecification can only happen in over-identified moment condition models. When the model is misspecified, the parameter of interest is set as the minimizer of the population GMM criterion function, which is referred to as the pseudo-true value. We assume that the pseudo-true value is unique. For more detail, see Kang and Lee (2024). 

For observations indexed by $i=1,...,n$, the one-step GMM estimator is defined as 
\begin{equation*}
	\widehat{\theta}_{1} = \argmin_{\theta} g_{n}(\theta)'W_n g_{n}(\theta),
\end{equation*}
where $g_{n}(\theta) = n^{-1}\sum_{i=1}^{n}g(X_{i},\theta), W_{n}$ is a positive definite weight matrix which takes the form of $n^{-1}\sum_{i=1}^{n}W(X_{i})$, and $W(X_{i})$ does not depend on any unknown parameter. We consider two common choices; 1) $W_n = I$, identity matrix, 2) $W(X_i) = 
(Z_{i}Z_{i}^{\prime})^{-1}$, where $Z_{i}$ is the instrument vectors.

The two-step efficient GMM is defined as
\begin{equation*}
	\widehat{\theta}_{2} = \argmin_{\theta} g_{n}(\theta)'\widehat{W}_{n}g_{n}(\theta),
\end{equation*}
where $\widehat{W}_{n} = \widehat{W}_{n}(\widehat{\theta}_{1})$ and 
\begin{equation*}
	\widehat{W}_{n}(\theta) = \left(\frac{1}{n}\sum_{i=1}^{n}g(X_{i},\theta)g(X_{i},\theta)'\right)^{-1}.
\end{equation*}

\section{Nonsmooth Location Model}

First, we consider a simple location model with i.i.d. data, following Hong and Li (2023). Specifically, the data is generated as:
\begin{equation*}
	y_{i} = \theta_{0} + \varepsilon_{i}, ~i=1,...,n,
\end{equation*}
where $\varepsilon_{i}\sim N(0,2^2)$ and $\theta_{0}=0$. The baseline moment function is defined as:
\begin{equation*}
	g_{1}(y_{i},\theta) = \left[\begin{array}{c}
		1(y_{i}\leq\theta)-\tau \\
		y_{i}-\theta
	\end{array}\right].
\end{equation*}
Since the distribution of $y_{i}$ is symmetric, the moment condition is correctly specified if $\tau=0.5$. If $\tau\neq0.5$ then the model is misspecified. To examine the effect of a misspecified nonlinear moment function (when $\tau \neq 0.5$), we also consider the following set of moment functions:
\begin{equation*}
		g_{2}(y_{i},\theta) = \left[\begin{array}{c}
		1(y_{i}\leq\theta)-\tau \\
		y_{i}-\theta\\
		(y_{i}-\theta)^2-4
	\end{array}\right].
\end{equation*}
The last equation of $g_{2}(y_{i},\theta)$ imposes a condition that the variance of $y_{i}$ is four, provided that $\theta$ is the mean. When $\tau\neq 0.5$, there is no parameter that satisfies the first two equations of the moment condition simultaneously, and thus, the moment condition is misspecified. In this case, the pseudo-true value differs from the mean, and the last equation does not hold either. 

Next, we add potentially misspecified parameter-free moment functions (if $\tau\neq0.5$):
\begin{equation*}
	g_{3}((y_{i},x_{i}')',\theta) = \left[\begin{array}{c}
		1(y_{i}\leq\theta)-\tau \\
		y_{i}-\theta\\
		(y_{i}-\theta)^2-4\\
		x_{i}-(0.5-\tau)
	\end{array}\right],
\end{equation*}
where $x_{i}\in\mathbb{R}^{5}$ is generated as:
\begin{equation*}
	\left(\begin{array}{c}
		\varepsilon_{i} \\
		x_{i}\\
	\end{array}\right)\sim N\left(\mathbf{0}_{6\times1},\left[\begin{array}{cccccc}
		1 & 0.5 & 0.4 & 0.3 & 0.2 & 0.1 \\
		0.5 & 1 & 0.5 & 0.4 & 0.3 & 0.2  \\
		0.4 & 0.5 & 1 & 0.5 & 0.4 & 0.3  \\
		0.3 & 0.4 & 0.5 & 1 & 0.5 & 0.4   \\
		0.2 & 0.3 & 0.4 & 0.5 & 1 & 0.5    \\
		0.1 & 0.2 & 0.3 & 0.4 & 0.5 & 1     \\
	\end{array}\right]\right).
\end{equation*}
Since $E[x_{i}]=0$, the last set of equations in $g_{3}((y_{i},x_{i}')',\theta)$ is misspecified if $\tau\neq0.5$.  

To compare the finite sample behavior of the GMM estimator using nonsmooth moments with those using smooth moments, we finally consider the following moment condition:
\begin{equation*}
	g_{4}((y_{i},x_{i}')',\theta) = \left[\begin{array}{c}
		y_{i}-\theta\\
		(y_{i}-\theta)^2-4\\
		x_{i}-(0.5-\tau)
	\end{array}\right],
\end{equation*}
where the variables are generated as specified above.

For each set of moment conditions, we generate $n=200,400,800, 1600, 3200, 6400$ observations and estimate $\theta$ by one-step GMM with the identify matrix as the weight matrix, and the two-step efficient GMM using the one-step GMM as the preliminary estimator. The number of Monte Carlo repetitions is 10,000. The simulation was conducted in MATLAB, where minimization was carried out using the \texttt{fminunc} function. The initial value of $\theta$ in the minimization problem was randomly generated from $U[-1,1]$. 

\begin{table}[btp]
\centering
\small
\begin{tabular}{cccccccc}
\toprule
$g_{1}$& & \multicolumn{3}{c}{$\widehat{\theta}_{1}$} &  \multicolumn{3}{c}{$\widehat{\theta}_{2}$}\\
 \cmidrule(lr){3-5} \cmidrule(l){6-8}
&$\tau$ & 0.1 & 0.3 & 0.5 & 0.1 & 0.3 & 0.5\\
\midrule
\multirow{6}{*}{$n$} & 200 & 0.0205& 0.0207& 0.0202 & 0.1115& 0.1187&0.0289\\
& 400 & 0.0104& 0.0103& 0.0100& 0.0787& 0.0886&0.0166 \\
& 800 & 0.0054& 0.0054& 0.0050 & 0.0471&0.0570 &0.0090\\
 & 1600 & 0.0027& 0.0027& 0.0025 & 0.0261& 0.0319&0.0046\\
 & 3200 & 0.0014& 0.0014& 0.0012 & 0.0152 & 0.0178 & 0.0024\\
& 6400 & 0.0008& 0.0007& 0.0006 & 0.0089& 0.0103 &0.0013\\
	\toprule
	$g_{2}$& & \multicolumn{3}{c}{$\widehat{\theta}_{1}$} &  \multicolumn{3}{c}{$\widehat{\theta}_{2}$}\\
	\cmidrule(lr){3-5} \cmidrule(l){6-8}
	&$\tau$ & 0.1 & 0.3 & 0.5 & 0.1 & 0.3 & 0.5\\
	\midrule
	\multirow{6}{*}{$n$} & 200 & 0.0367& 0.0383& 0.0351 & 0.1097& 0.1199&0.0315\\
	& 400 & 0.0166& 0.0162& 0.0132& 0.0795& 0.0893&0.0168 \\
	& 800 & 0.0083& 0.0071& 0.0052 & 0.0487&0.0570 &0.0089\\
	& 1600 & 0.0041& 0.0034& 0.0025 & 0.0275& 0.0317&0.0045\\
	& 3200 & 0.0019& 0.0017& 0.0012 & 0.0155 & 0.0174 & 0.0024\\
	& 6400 & 0.0010& 0.0008& 0.0006 & 0.0090& 0.0102 &0.0013\\
	\toprule
	$g_{3}$& & \multicolumn{3}{c}{$\widehat{\theta}_{1}$} &  \multicolumn{3}{c}{$\widehat{\theta}_{2}$}\\
	\cmidrule(lr){3-5} \cmidrule(l){6-8}
	&$\tau$ & 0.1 & 0.3 & 0.5 & 0.1 & 0.3 & 0.5\\
	\midrule
	\multirow{6}{*}{$n$} & 200 & 0.0372& 0.0386& 0.0357 & 0.1493& 0.1268&0.0230\\
	& 400 & 0.0171& 0.0166& 0.0135& 0.1202& 0.0926&0.0115 \\
	& 800 & 0.0084& 0.0072& 0.0053 & 0.0911&0.0570 &0.0055\\
	& 1600 & 0.0042& 0.0035& 0.0025 & 0.0681& 0.0274&0.0029\\
	& 3200 & 0.0020& 0.0017& 0.0012 & 0.0445 & 0.0131 & 0.0015\\
	& 6400 & 0.0010& 0.0008& 0.0006 & 0.0235& 0.0083 &0.0008\\
	\toprule
$g_{4}$ & & \multicolumn{3}{c}{$\widehat{\theta}_{1}$} &  \multicolumn{3}{c}{$\widehat{\theta}_{2}$}\\
\cmidrule(lr){3-5} \cmidrule(l){6-8}
&$\tau$ & 0.1 & 0.3 & 0.5 & 0.1 & 0.3 & 0.5\\
\midrule
\multirow{6}{*}{$n$} & 200 & 0.0367& 0.0367& 0.0367 & 0.0158& 0.0154&0.0154\\
& 400 & 0.0138& 0.0138& 0.0138& 0.0078& 0.0075&0.0074 \\
& 800 & 0.0053& 0.0053& 0.0053 & 0.0038&0.0037 &0.0036\\
& 1600 & 0.0025& 0.0025& 0.0025 & 0.0019& 0.0018&0.0018\\
& 3200 & 0.0012& 0.0012& 0.0012 & 0.0009 & 0.0009 & 0.0009\\
& 6400 & 0.0006& 0.0006& 0.0006 & 0.0005& 0.0005 &0.0004\\
	\bottomrule
\end{tabular}
\caption{Variance of the GMM estimator }
\label{T1}
\end{table}

Table \ref{T1} shows the variance of the one-step and two-step GMM estimators under each set of the moment conditions with increasing sample size. Note that the variance of the $\sqrt{n}$-convergent estimator decays faster ($n^{-1}$) than the $n^{1/3}$-convergent estimator ($n^{-2/3}$). 

When the model is correctly specified ($\tau=0.5$), the variance of the estimator decreases inversely proportional to the sample size, consistent with theoretical predictions. This holds regardless of the moment condition used and whether the estimator is one-step or two-step GMM.

In contrast, under misspecification (\(\tau \neq 0.5\)), it is expected that the variance of the estimator does not decrease at the same rate with increasing sample size. Surprisingly, the simulation results show otherwise. For the one-step GMM estimator \(\widehat{\theta}_1\) with the identity matrix as the weight matrix, the variance decreases at the same rate as under correct specification. This finding differs from the theoretical results of Hong and Li (2023), who established a cubic-root convergence rate for the GMM estimator when a fixed weight matrix is used. In particular, Hong and Li considered a similar simulation setting (except for the variance of \(\varepsilon_i\)) in Section 5.1 to demonstrate the superior finite-sample coverage of their adaptive bootstrap confidence intervals (CIs) compared to standard bootstrap CIs. The results in Table \ref{T1} suggest that the improved performance of the adaptive bootstrap reported in their Table 1 may stem from factors other than the convergence rate, such as the recentering procedure when calculating the bootstrap GMM estimator.

On the other hand, the variance of the two-step efficient GMM estimator $\widehat{\theta}_{2}$ decreases at a slower rate than $n^{-1}$, approximately $n^{-2/3}$. Adding a nonlinear moment condition ($g_1$ versus $g_2$) does not significantly affect the convergence rate. However, adding more misspecified moment conditions ($g_2$ versus $g_3$) generally slows the convergence rate further.

\begin{figure}[btp]
	\centering
	\includegraphics[width=1\linewidth]{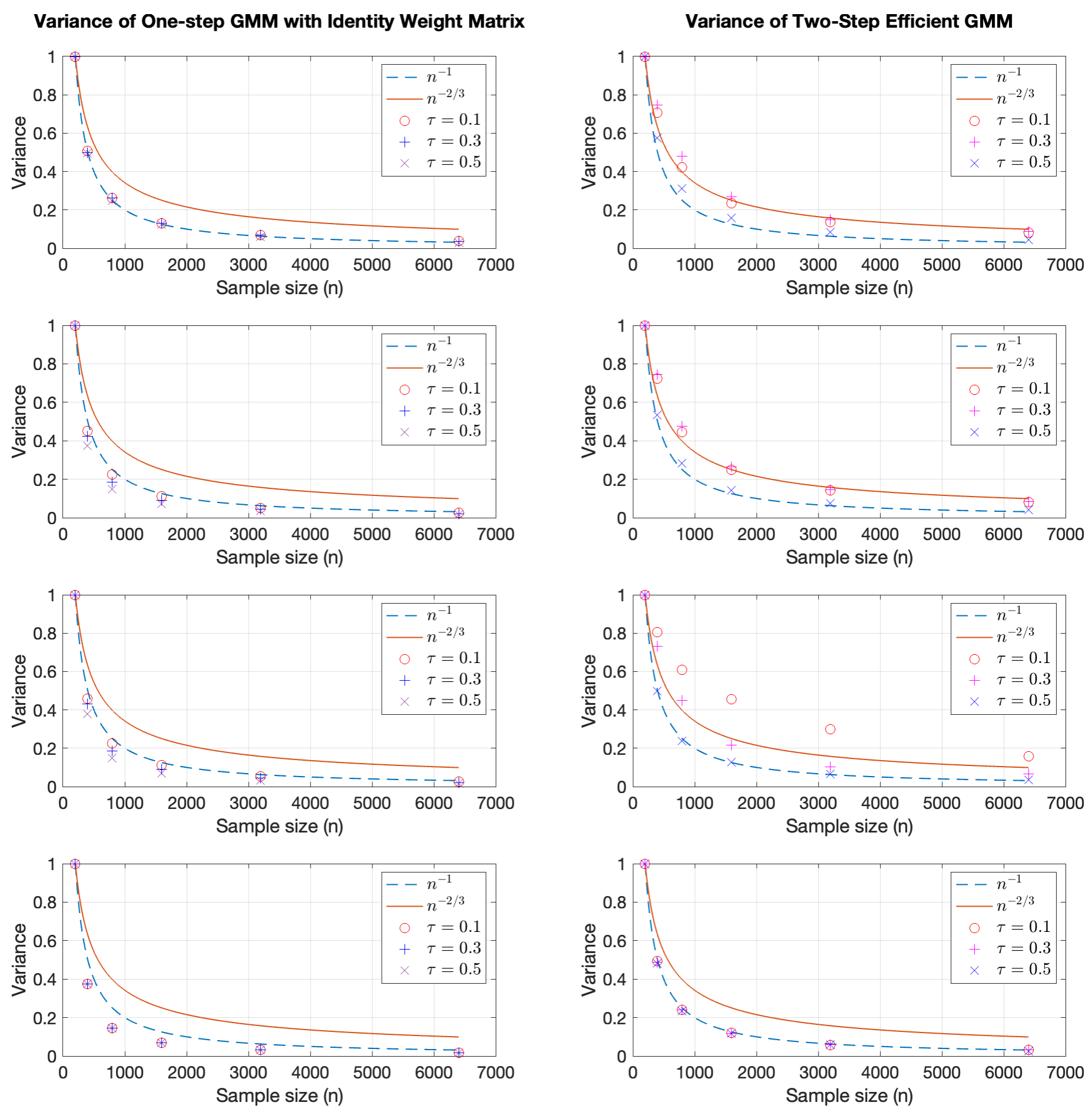}
	\caption{Variance Decay Rates of the one-step and two-step GMM Estimators.\\First row ($g_1$), Second row ($g_2$), Third row ($g_3$), Fourth row ($g_4$)}
	\label{fig1}
\end{figure}

To illustrate the difference in the variance decay rates, we normalize the variance relative to its value at  $n = 200$  for each $\tau$, $n$, and set of moment conditions. The results are presented in Figure \ref{fig1}. The left column shows the variance of the one-step GMM estimator with the identity weight matrix, while the right column displays the variance of the two-step efficient GMM estimator. Each row corresponds to one of the sets of moment conditions:  $g_1$, $g_2$, $g_3$, and $g_4$.

The variance of the one-step GMM estimator with identity matrix decays at approximately an $n^{-1}$ rate, even under misspecification ($\tau$ = 0.1, 0.3). In contrast, the variance of the two-step efficient GMM estimator with nonsmooth moments decays significantly more slowly under misspecification.

\section{Quantile Regression with Endogeneity} 

The IV quantile regression was developed by Chernozhukov and Hansen (2005). They consider a model of quantile treatment effects (QTE) in the presence of endogeneity and derive the moment conditions necessary for the identification of QTE without imposing functional form assumptions. This approach provides both economic and causal justification for estimation based on these restrictions.

Consider a linear quantile regression model with endogeneity characterized by the structural equation
\begin{align}
	\label{str}
    Y &= D'\alpha(U)+X'\beta(U),~~ U|X,Z \sim \text{Uniform}(0,1),  \\
    \tau &\mapsto D'\alpha(U)+X'\beta(U) ~~\text{strictly increasing in}~ \tau
    \notag
\end{align}
where $Y$ is the scalar outcome variable of interest, $U$ is an unobserved scalar random variable, and $X$ is a vector of included control variables. The covariates $D$ may not be independent of $U$. We assume that there is a vector of instrumental variables, denoted by $Z$, which is excluded from equation \eqref{str} but affects the endogenous variables $D$, with $\text{dim}(Z)\geq \text{dim}(D)$. 

Under these assumptions, it follows that, for $\tau \in (0, 1)$,
\begin{equation}
	P[Y \leq D'\alpha(\tau)+X'\beta(\tau) \mid X, Z] = P[U \leq \tau \mid X, Z] = \tau.
	\label{ivqr1}
\end{equation}
In this model, $\alpha(\tau)$ and $\beta(\tau)$ capture the effects of the covariates $D$ and $X$ on the outcome variable for an individual whose unobserved heterogeneity $U$ is fixed at $U = \tau$. By the definition of probability and the law of iterated expectation, \eqref{ivqr1} implies the following unconditional moment condition:
\begin{equation}
	E([\tau - 1(Y \leq D'\alpha(\tau)+X'\beta(\tau)]\Psi) = 0,
	\label{ivqr2}
\end{equation}
where $1(\cdot)$ is an indicator function and $\Psi = (X', Z')'$ is a vector of instruments and covariates.

Given these moment conditions, GMM estimation is appropriate. However, since the moment function is discontinuous in the parameters of interest, applying conventional minimization techniques is challenging. Consequently, various methods for IVQR estimation have been developed subsequently.

Chernozhukov and Hansen (2006, 2008) proposed a method called Inverse Quantile Regression (IQR), which estimates IVQR using the traditional quantile regression approach combined with a grid search algorithm. The IQR method is carried out as follows. Define the quantile regression objective function as:
\begin{align*}
	Q_n(\tau, \alpha, \beta, \gamma) := \frac{1}{n}\sum_{i=1}^{n}\rho_\tau(Y_i - D_i'\alpha - X_i'\beta - \widehat{\Phi}_i'\gamma),
\end{align*}
where $\widehat{\Phi}_i \equiv f(X_i, Z_i)$ is a $\text{dim}(\alpha) \times 1$ vector of (transformations of) instruments, and the check function $\rho_\tau$ is defined as $\rho_\tau(u) = u(\tau - 1\{u < 0\})$ for $u \in \mathbb{R}$.

For a given value of the structural parameter $\alpha$, the ordinary quantile regression is run using the objective function above, and a value of $\alpha$ is obtained that minimizes the coefficient on the instrumental variable, $\widehat{\gamma}(\alpha, \tau)$, as close to 0 as possible. Formally, let:
\begin{align*}
	\widehat{\alpha}(\tau) \equiv \arginf_{\alpha \in \mathcal{A}} \widehat{\gamma}(\alpha, \tau)'A\widehat{\gamma}(\alpha, \tau), \quad \text{where} \quad \left(\widehat{\beta}(\alpha, \tau), \widehat{\gamma}(\alpha, \tau)\right) = \arginf_{(\beta, \gamma) \in \mathcal{B} \times \mathcal{G}} Q_n(\tau, \alpha, \beta, \gamma),
\end{align*}
where $A$ is any positive definite matrix, and $\mathcal{A}$, $\mathcal{B}$, and $\mathcal{G}$ are compact parameter spaces. The IQR estimator, denoted by $\widehat{\theta}_{\text{iqr}}(\tau)$, is then defined as:
\[
\widehat{\theta}_{\text{iqr}}(\tau) = \left( \widehat{\alpha}(\tau), \widehat{\beta}(\widehat{\alpha}(\tau), \tau)\right).
\]

Kaplan and Sun (2017) proposed another method, smoothed IVQR referred to as SIVQR. This approach smooths the underlying moment condition by applying a kernel to the indicator function in \eqref{ivqr2}. Replacing $1(\cdot)$ with a similar but continuously differentiable function $\tilde{1}(\cdot)$ enables GMM estimation based on smooth moment conditions. When the model is overidentified with $\text{dim}(Z) > \text{dim}(D)$, they transform the original moment condition (11) into:
\[
E\left([\tau - \tilde{1}(Y \leq D'\alpha(\tau)+X'\beta(\tau))]\widetilde{\Phi}\right) = 0,
\]
where $\widetilde{\Phi} = (\hat{D}', X')'$ and $\hat{D}$ is a linear transformation of $X$ and $Z$ that has the same dimension as $D$. This transformation results in an exactly identified model with the transformed instrument vector $\widetilde{\Phi}$. Both IQR and SIVQR use the transformed instruments obtained from the least squares projection of $D$ onto $Z$ and $X$ in practice. See Secion \ref{exact} and Kaplan and Sun (2017) for further details.

Machado and Santos Silva (2019) proposed an estimator for conditional quantiles by combining estimates of the location and scale functions, referred to as the Method of Moments-Quantile Regression (MM-QR). The conditional location-scale model is given by:
\[
Y = X'\beta + \sigma(X'\gamma)U,
\]
where $Y$ is the scalar outcome variable, $X$ includes the endogenous variable $D$ and other exogenous covariates, and $\sigma(\cdot)$ is a known function.

Based on normalization of the unobserved random variable $U$, Machado and Santos Silva (2019) use the moment conditions $E[ZU] = 0$ and $E[Z(|U| - 1)] = 0$ with instruments $Z$ to obtain consistent estimates of $\beta$ and $\gamma$ under very general conditions by applying GMM.

Given the estimates of $\beta$ and $\gamma$, $q(\tau)$ can be estimated using the following moment condition:
\[
E\left[\tau - 1\left(\frac{Y - X'\beta}{\sigma(X'\gamma)} \leq q(\tau)\right)\right] = 0,
\]
where traditional quantile regression can be applied to the estimated residuals. By combining $\widehat{\beta}$, $\widehat{\gamma}$, and $\widehat{q}(\tau)$, the estimates of the desired regression quantile coefficient can be obtained.

The estimators obtained through these methods are not IVQR estimates within the classical GMM framework, which directly uses the moment conditions in equation \eqref{ivqr2} for GMM estimation. We investigate the convergence rates of the estimators obtained through these various estimation methods compared to the GMM estimator under misspecification.

\subsection{Simulation Results for IVQR} 
The data generating process for IVQR estimation, as given in Hong and Li (2023), is as follows. For $\alpha_0 = \beta_0= 1$,
\begin{align*}
	Y_{i} &= \alpha_{0} + \beta_0 D_{i} + u_{i},~~\left(\begin{array}{c}
		u_{i} \\
		D_{i} \\
		W_{i} \\
	\end{array}\right) \sim N\left(\left(\begin{array}{c}
		0 \\ 
		0 \\
		0 \\
	\end{array}\right),\left(\begin{array}{ccc}
		1 & 0 & \delta \\
		0 & 1 & 0.5 \\
		\delta & 0.5 & 1 \\
	\end{array}\right)\right).
\end{align*}
It follows that:
\begin{align*}
	y_i|D_i,W_{i} \sim N &\Bigg(\alpha_0+\beta_0D_i +\left(-\frac{2}{3}D_i+\frac{4}{3}W_i\right)\delta, 1- \frac{4}{3}\delta^2\Bigg) .
\end{align*}     
Considering median regression, the population moments for $z_i = (1 ~~ D_i ~~ W_i)'$ and $\theta=(\alpha,\beta)'$ are given by
\begin{align*}
	\pi(\theta) &= E\left[\left(\frac{1}{2} - 1(y_i \leq \alpha + \beta D_i)\right)z_i\right] \\
	&= E\left[\left(\frac{1}{2} - F_{y|D,W}(\alpha + \beta D_i)\right)z_i\right] \\
	&= E\left[\left(\frac{1}{2} - \Phi\left( \frac{\alpha - \alpha_o + (\beta - \beta_0)D_i + (\frac{2}{3}D_i - \frac{4}{3}W_i)\delta}{\sqrt{1-\frac{4}{3}\delta^2}}\right)\right)z_i\right].
\end{align*}     
At the true parameter values, the population moments become:
\begin{align*}
	\pi(\theta_0)
	&= E\left[\left(\frac{1}{2} - \Phi\left( \frac{(\frac{2}{3}D_i - \frac{4}{3}W_i)\delta}{\sqrt{1-\frac{4}{3}\delta^2}}\right)\right)z_i\right] 
\end{align*}  
Here, $\Phi(\cdot)$ is the cumulative standard normal distribution function. This model is correctly specified for median regression when $\delta = 0$. However, when $\delta \neq 0$, the model becomes misspecified.

We generated $n = 200, 400, 800, 1600, 3200, 6400$ observations and varied the degree of misspecification by setting $\delta = 0, 0.1, 0.2, 0.4, 0.6$ to estimate $\alpha_0$ and $\beta_0$. The exact computation of the GMM estimator for IVQR models follows a mixed-integer quadratic programming approach, as proposed by Chen and Lee (2018). Two types of one-step GMM estimators are considered:
\begin{enumerate}
	\item Fixed Weight: Uses the identity matrix as the weight matrix.
	\item Estimated Weight: Employs the following weight matrix:
	\[
	\widehat{W} = \left[\tau(1 - \tau)n^{-1}\sum z_i z_i'\right]^{-1}.
	\]
\end{enumerate}

The simulation was conducted in MATLAB using Gurobi as the numerical solver. Both time and gap were set to zero to ensure full convergence. The number of Monte Carlo repetitions was 1,000.

Table \ref{T3} reports the variance of the GMM estimators in median regression as the sample size increases. Results are presented only for the coefficient of the variable of interest, $D_i$. For both one-step GMM estimators with fixed weight ($\widehat{\beta}_{\text{fixed}}$) and estimated weight ($\widehat{\beta}_{\text{est}}$), the variance decreases at the rate of $n^{-1}$ when the model is correctly specified ($\delta = 0$).

\begin{table}[btp]
	\centering
	\small
	\begin{tabular}{ccccccc}
		\toprule
		\multicolumn{7}{c}{$\widehat{\beta}_{\text{fixed}}$}\\
		\cmidrule(lr){3-7}
		&$\delta$ & 0 & 0.1 & 0.2 & 0.4 & 0.6\\
		\midrule
		\multirow{6}{*}{$n$} & 200   & 0.00884 & 0.00966 & 0.01039 & 0.01472 & 0.02122 \\
		&400   & 0.00417 & 0.00452 & 0.00569 & 0.00803 & 0.00982 \\
		&800   & 0.00221 & 0.00257 & 0.00279 & 0.00415 & 0.00445 \\
		&1600  & 0.00110 & 0.00129 & 0.00150 & 0.00215 & 0.00095 \\
		&3200  & 0.00052 & 0.00061 & 0.00075 & 0.00065 & 0.00035 \\
		&6400  & 0.00027 & 0.00033 & 0.00043 & 0.00018 & 0.00016 \\
		
		\toprule
		\multicolumn{7}{c}{$\widehat{\beta}_{\text{est}}$}\\
		\cmidrule(lr){3-7}
		&$\delta$ & 0 & 0.1 & 0.2 & 0.4 & 0.6\\
		\midrule
		\multirow{6}{*}{$n$} & 200   & 0.00871 & 0.00935 & 0.01065 & 0.01817 & 0.03065 \\
		&400   & 0.00409 & 0.00459 & 0.00580 & 0.01040 & 0.01887 \\
		&800   & 0.00215 & 0.00247 & 0.00346 & 0.00626 & 0.01177 \\
		&1600  & 0.00101 & 0.00131 & 0.00183 & 0.00364 & 0.00731 \\
		&3200  & 0.00053 & 0.00069 & 0.00109 & 0.00203 & 0.00412 \\
		&6400  & 0.00025 & 0.00039 & 0.00061 & 0.00137 & 0.00252 \\
		\bottomrule
	\end{tabular}
	\caption{Variance of the IVQR-GMM Estimator }
	\label{T3}
\end{table}

In contrast, under misspecification ($\delta = 0.1, 0.2, 0.4, 0.6$), the convergence rate of $\widehat{\beta}_{\text{est}}$ approaches $n^{-2/3}$ as the degree of misspecification increases. However, regardless of the value of $\delta$, the convergence rate of $\widehat{\beta}_{\text{fixed}}$ consistently remains at $n^{-1/2}$. These results align with the simulation findings from the previously discussed location model.

\begin{figure}[t]
	\includegraphics[width=1\linewidth]{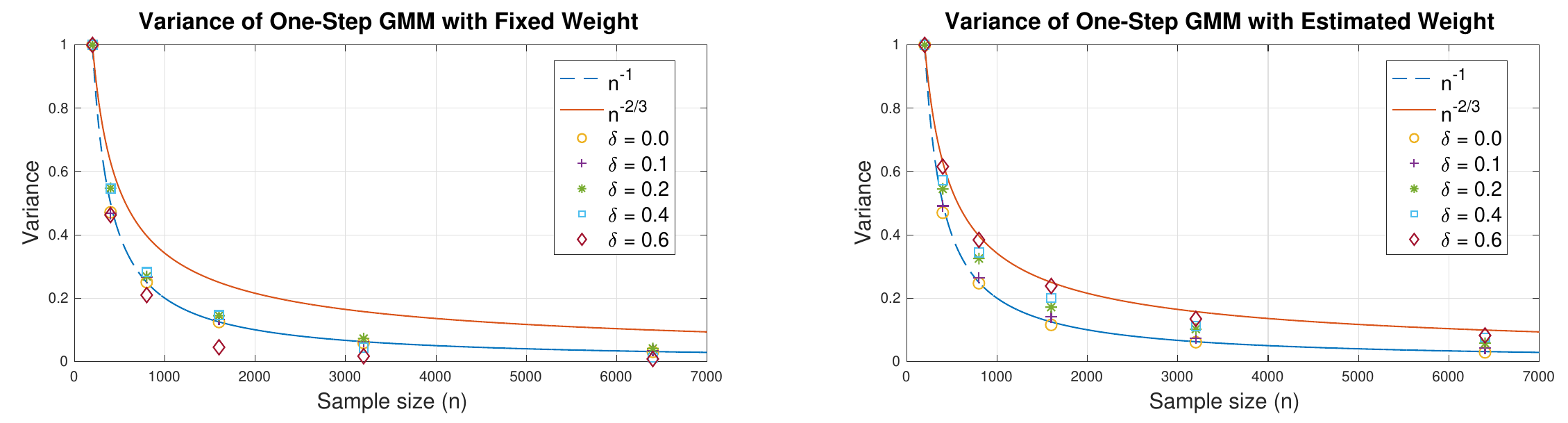}
	\caption{Variance Decay Rates of the IVQR One-step GMM Estimators}
	\label{fig3}
\end{figure}

This phenomenon is more clearly demonstrated in Figure \ref{fig3}. To highlight the differences in variance decay rates, we normalize the variance relative to its value at $n = 200$ for each sample size and each $\delta$. Under misspecification, the variance of the one-step GMM estimator decays at approximately an $n^{-1}$ rate. However, the convergence rate of the efficient GMM estimator gradually decreases as $\delta$ increases, eventually reaching the $n^{-2/3}$ rate only when $\delta$ is sufficiently large, indicating strong misspecification.

Additionally, we calculated the MM-QR estimator proposed by Machado and Santos Silva (2019) using the same DGP, which applies GMM estimation with a directionally differentiable moment condition. In STATA, the \texttt{ivqreg2} command provides an accessible way to compute MM-QR, using the identity function as the scale function. Similar to the previous analysis, the simulation was conducted with $n = 200, 400, 800, 1600, 3200, 6400$ and $\delta = 0, 0.1, 0.2, 0.4, 0.6$. Each simulation was repeated 1,000 times. For a fair comparison, we reported the estimation results for $\beta_0$ in the median regression.

\begin{table}[t]
	\centering
	\small
	\begin{tabular}{ccccccc}
		\toprule
		\multicolumn{7}{c}{$\widetilde{\beta}_{1}$ (One-step)}\\
		\cmidrule(lr){3-7}
		&$\delta$ & 0 & 0.1 & 0.2 & 0.4 & 0.6\\
		\midrule
		\multirow{6}{*}{$n$} & 200   & 0.00583 & 0.00595 & 0.00630 & 0.00792 & 0.01183 \\
		& 400   & 0.00279 & 0.00286 & 0.00303 & 0.00380 & 0.00546 \\
		& 800   & 0.00139 & 0.00141 & 0.00149 & 0.00183 & 0.00259 \\
		& 1600  & 0.00074 & 0.00075 & 0.00079 & 0.00097 & 0.00138 \\
		& 3200  & 0.00034 & 0.00034 & 0.00036 & 0.00046 & 0.00067 \\
		& 6400  & 0.00018 & 0.00019 & 0.00019 & 0.00024 & 0.00033 \\
		
		\toprule
		\multicolumn{7}{c}{$\widetilde{\beta}_{2}$ (Two-step)}\\
		\cmidrule(lr){3-7}
		&$\delta$ & 0 & 0.1 & 0.2 & 0.4 & 0.6\\
		\midrule
		\multirow{6}{*}{$n$} & 200   & 0.00526 & 0.00530 & 0.00562 & 0.00765 & 0.01233 \\
		& 400   & 0.00255 & 0.00261 & 0.00281 & 0.00382 & 0.00603 \\
		& 800   & 0.00124 & 0.00128 & 0.00139 & 0.00189 & 0.00289 \\
		& 1600  & 0.00067 & 0.00068 & 0.00072 & 0.00097 & 0.00152 \\
		& 3200  & 0.00031 & 0.00032 & 0.00035 & 0.00050 & 0.00079 \\
		& 6400  & 0.00016 & 0.00016 & 0.00017 & 0.00024 & 0.00040 \\
		\bottomrule
	\end{tabular}
	\caption{Variance of the MMQR Estimator}
	\label{T4}
\end{table}

Table \ref{T4} presents the variance of the one-step MMQR estimator $\widetilde{\beta}_1$ and the two-step MMQR estimator $\widetilde{\beta}_2$ as the sample size increases. The variance of both $\widetilde{\beta}_1$ and $\widetilde{\beta}_2$ decays at approximately an $n^{-1}$ rate as $n$ increases, for any value of $\delta$. According to Kang and Lee (2024), when GMM estimation is conducted using a nonsmooth but directionally differentiable moment condition, the convergence rate of the estimator should be $n^{-1/2}$, regardless of whether the model is correctly specified or misspecified. The simulation results align with this theoretical expectation.

\begin{figure}[t]
	\includegraphics[width=1\linewidth]{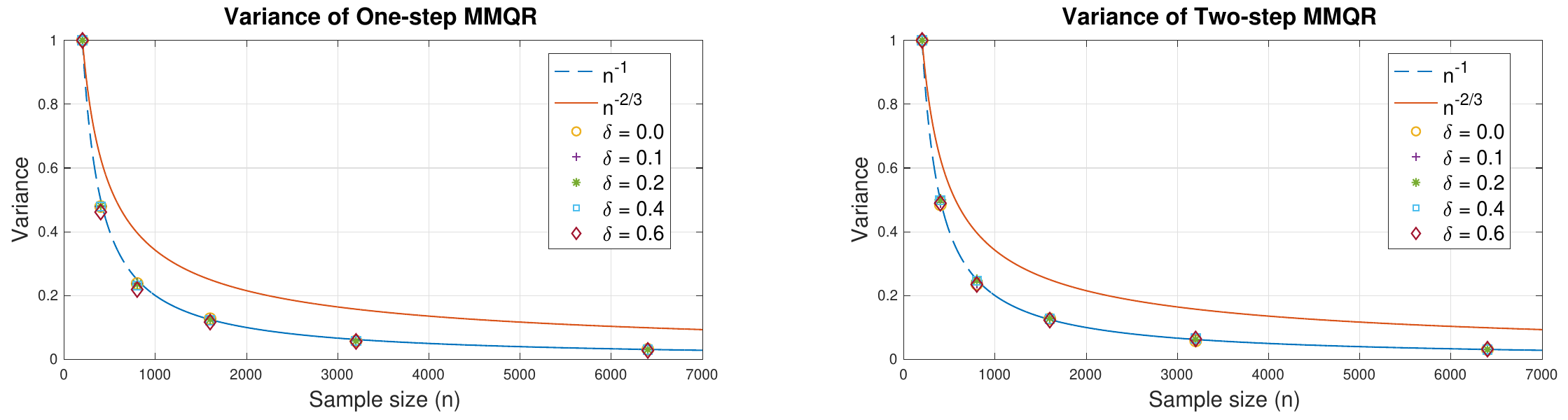}
	\caption{Variance Decay Rates of the One-step and Two-step MMQR Estimators}
	\label{fig4}
\end{figure}

Figure \ref{fig4} illustrates the variance of the MMQR estimator, normalized to its value at $n = 200$. Unlike the results of exact GMM estimation (Chen and Lee, 2018), the MMQR estimator demonstrates that the convergence rate of the variance is approximately $n^{-1}$, regardless of whether fixed or estimated weights are used, or the value of $\delta$.

\subsection{Transforming over-identified moment conditions into exactly identified ones} 
\label{exact}
Chernozhukov and Hansen (2006) and Kaplan and Sun (2017) propose a transformation of over-identified moment conditions into exactly-identified moment conditions in quantile regression models with endogeneity. While this transformation may offer computational advantages, it may also obscure potential misspecification in the original over-identified moment condition model.

For illustration, consider the following simple linear model with no constant and endogeneity:
\begin{equation*}
	Y = D\beta_0 + e,~~E[De]\neq0.
\end{equation*}
Suppose that there are two instruments, $Z_1$ and $Z_2$, where $E[Z_1 e]=0$ but $E[Z_2 e]\neq0$. In other words, $Z_1$ is a valid instrument, but $Z_2$ is not. Both satisfy the relevance condition: $E[Z_{1}D]\neq0$ and $E[Z_{2}D]\neq0$. Then the moment condition given by
\begin{equation*}
	E\left[\begin{array}{c}
		Z_{1}(Y-D\beta)\\
		Z_{2}(Y-D\beta)
	\end{array}\right]
\end{equation*}
is misspecified because there is no $\beta$ that simultaneously satisfies the moment condition. The IV estimands are 
\begin{equation*}
	\frac{E[Z_{1}Y]}{E[Z_{1}D]} = \beta_{0},~~\frac{E[Z_{2}Y]}{E[Z_{2}D]} = \beta^{*},
\end{equation*}
and $\beta_{0}\neq\beta^{*}$. Now consider a transformation of the original over-identified moment condition into the exactly identified moment condition via a $1\times 2$ matrix $\Pi = \left(\pi_{1},~\pi_{2}\right)$:
\begin{equation}
	E\left[\Pi\left(\begin{array}{c}
		Z_{1}(Y-D\beta)\\
		Z_{2}(Y-D\beta)
	\end{array}\right)\right] = E[\pi_{1}Z_{1}(Y-D\beta) + \pi_{2}Z_{2}(Y-D\beta)].
	\label{exactMC}
\end{equation}
By solving this moment condition, we find that \eqref{exactMC} equals to zero at 
\begin{equation*}
	\overline{\beta} = \frac{\pi_{1}E[Z_{1}D]}{\pi_{1}E[Z_{1}D] + \pi_{2}E[Z_{2}D]} \beta_{0} + \frac{\pi_{2}E[Z_{2}D]}{\pi_{1}E[Z_{1}D] + \pi_{2}E[Z_{2}D]} \beta^{*}.
\end{equation*}
Provided that $\pi_{2}\neq0$, $\overline{\beta}\neq\beta_{0}$. Therefore, the exactly identified moment condition holds at a parameter value which is different from the true value. But the standard specification tests such as the J test cannot be applied to the exactly identified moment condition. 

By transforming the instruments using the least squares projection of $D$ onto $Z_1$ and $Z_2$, an over-identified model can be converted into a just-identified model. Since a $\beta$ that satisfies such moment conditions always exists, moment misspecification, which can occur in over-identified models, cannot arise. 

Moreover, the IQR method proposed by Chernozhukov and Hansen (2006) is not a GMM estimator in finite samples\footnote{However, it is asymptotically a GMM estimator.} because it relies on grid search for estimation. Similarly, the SIVQR method proposed by Kaplan and Sun (2017) is not a GMM estimator with nonsmooth moments because it smooths the indicator function using a kernel. Thus, estimating IVQR using the methods of Chernozhukov and Hansen (2006) and Kaplan and Sun (2017) does not align with the misspecification scenario considered in Hong and Li (2023). 

This implies that both estimators obtained using these two methods exhibit the standard $\sqrt{n}$-consistency, even under misspecification.

\subsection{Simulation Results for IQR and SIVQR}
To examine the convergence rate of each estimator when the IVQR model is estimated using the methods proposed by Chernozhukov and Hansen (2006) and Kaplan and Sun (2017), we considered the following DGP from Kang and Lee (2024):
\begin{align*}
	y_i &= -1 + D_i + \delta(z_{1i} - z_{2i}) + (1 + D_i)\epsilon_i, \\
	D_i &= \Phi(z_{1i} + z_{2i} + z_{3i} + v_i),
\end{align*}
where $(z_{1i},z_{2i},z_{3i})\sim N(0,I_{3})$ and $(\epsilon_{i},v_{i})\sim N(0,I_{2})$. Note that $\delta$ is a parameter that controls misspecification, where $\delta=0$ represents a correctly specified model, while any other value indicates a misspecified model.

If $\delta=0$, the above model can be rewritten using the Skorohod representation as follows:
\begin{equation}
	y_i = \alpha_0(U) + \beta_0(U)D_i,
	\label{ivqr}
\end{equation}
where $U = F_{\epsilon}(\epsilon)$ with $F_{\epsilon}$ being the cumulative distribution function of the unobservable $\epsilon$, and
\begin{align*}
	\alpha_0(\tau) = -1 + F_{\epsilon}^{-1}(\tau), \quad \beta_0(\tau) = 1 + F_{\epsilon}^{-1}(\tau). 
\end{align*}

We generated $n = 200, 400, 800, 1600, 3200, 6400$ observations and considered $\tau = 0.25, 0.5, 0.75$. The number of Monte Carlo repetitions is 10,000. The simulation was conducted in STATA. Estimation using the IQR method of Chernozhukov and Hansen (2006) was performed with the \texttt{ivqregress iqr} command, while the SIVQR method of Kaplan and Sun (2017) was implemented using the \texttt{ivqregress smooth} command. Both estimators estimate $\alpha_{0}(\tau)$ and $\beta_{0}(\tau)$ in \eqref{ivqr}.

\begin{table}[bt]
\centering
\small
\begin{tabular}{cccccccccc}
\toprule
$\tau=0.25$& & \multicolumn{4}{c}{$\widehat{\beta}_{iqr}$} &  \multicolumn{4}{c}{$\widehat{\beta}_{sivqr}$}\\
 \cmidrule(lr){3-6} \cmidrule(l){7-10}
&$\delta$ & 0 & 0.1 & 0.2 & 0.3 & 0 & 0.1 & 0.2 & 0.3\\
\midrule
\multirow{6}{*}{$n$} & 200 & 0.2125 & 0.2149 & 0.2242 & 0.2374 & 0.1758 & 0.1790 & 0.1864 & 0.1982  \\
& 400 & 0.1092 & 0.1100 & 0.1140 & 0.1210 & 0.0924 & 0.0935 & 0.0970 & 0.1032  \\
& 800 & 0.0534 & 0.0543 & 0.0564 & 0.0599 & 0.0464 & 0.0470 & 0.0488 & 0.0518  \\
& 1600 & 0.0266 & 0.0271 & 0.0284 & 0.0301 & 0.0237 & 0.0241 & 0.0250 & 0.0266  \\
& 3200 & 0.0135 & 0.0137 & 0.0142 & 0.0150 & 0.0122 & 0.0123 & 0.0127 & 0.0135  \\
& 6400 & 0.0069 & 0.0069 & 0.0072 & 0.0076 & 0.0062 & 0.0063 & 0.0066 & 0.0070  \\
\toprule
$\tau=0.5$& & \multicolumn{4}{c}{$\widehat{\beta}_{iqr}$} &  \multicolumn{4}{c}{$\widehat{\beta}_{sivqr}$}\\
 \cmidrule(lr){3-6} \cmidrule(l){7-10}
&$\delta$ & 0 & 0.1 & 0.2 & 0.3 & 0 & 0.1 & 0.2 & 0.3\\
	\midrule
	\multirow{6}{*}{$n$} & 200 & 0.1827 & 0.1847 & 0.1904 & 0.2034 & 0.1564 & 0.1580 & 0.1635 & 0.1733  \\
& 400 & 0.0924 & 0.0934 & 0.0972 & 0.1053 & 0.0799 & 0.0816 & 0.0854 & 0.0914  \\
& 800 & 0.0449 & 0.0453 & 0.0469 & 0.0506 & 0.0395 & 0.0398 & 0.0415 & 0.0445  \\
& 1600 & 0.0223 & 0.0228 & 0.0238 & 0.0253 & 0.0201 & 0.0204 & 0.0212 & 0.0226  \\
& 3200 & 0.0114 & 0.0117 & 0.0122 & 0.0131 & 0.0104 & 0.0106 & 0.0111 & 0.0118  \\
& 6400 & 0.0058 & 0.0058 & 0.0061 & 0.0065 & 0.0053 & 0.0054 & 0.0056 & 0.0059  \\
	\toprule
	$\tau=0.75$& & \multicolumn{4}{c}{$\widehat{\beta}_{iqr}$} &  \multicolumn{4}{c}{$\widehat{\beta}_{sivqr}$}\\
	\cmidrule(lr){3-6} \cmidrule(l){7-10}
&$\delta$ & 0 & 0.1 & 0.2 & 0.3 & 0 & 0.1 & 0.2 & 0.3\\
	\midrule
	\multirow{6}{*}{$n$} & 200 & 0.2172 & 0.2190 & 0.2251 & 0.2394 & 0.1789 & 0.1811 & 0.1875 & 0.1983  \\
& 400 & 0.1054 & 0.1067 & 0.1122 & 0.1185 & 0.0901 & 0.0915 & 0.0956 & 0.1019  \\
& 800 & 0.0535 & 0.0546 & 0.0560 & 0.0589 & 0.0463 & 0.0470 & 0.0486 & 0.0514  \\
& 1600 & 0.0267 & 0.0271 & 0.0281 & 0.0297 & 0.0236 & 0.0238 & 0.0247 & 0.0263  \\
& 3200 & 0.0136 & 0.0138 & 0.0142 & 0.0149 & 0.0122 & 0.0124 & 0.0128 & 0.0136  \\
& 6400 & 0.0067 & 0.0068 & 0.0070 & 0.0074 & 0.0061 & 0.0061 & 0.0063 & 0.0067  \\
	\bottomrule
\end{tabular}
\caption{Variance of the IQR and SIVQR Estimators for $\beta_0(\tau)$}
\label{T2}
\end{table}

In Table \ref{T2}, the variance of IQR and SIVQR estimator decreases at a rate of $n^{-1}$, regardless of the value of $\delta$. As explained in Section \ref{exact}, this is because both IQR and SIVQR transform the over-identified model into a just-identified model through linear projection, which prevents the slowdown in the convergence rate of the estimator under moment misspecification.

\begin{figure}[btp]
    \includegraphics[width=1\linewidth]{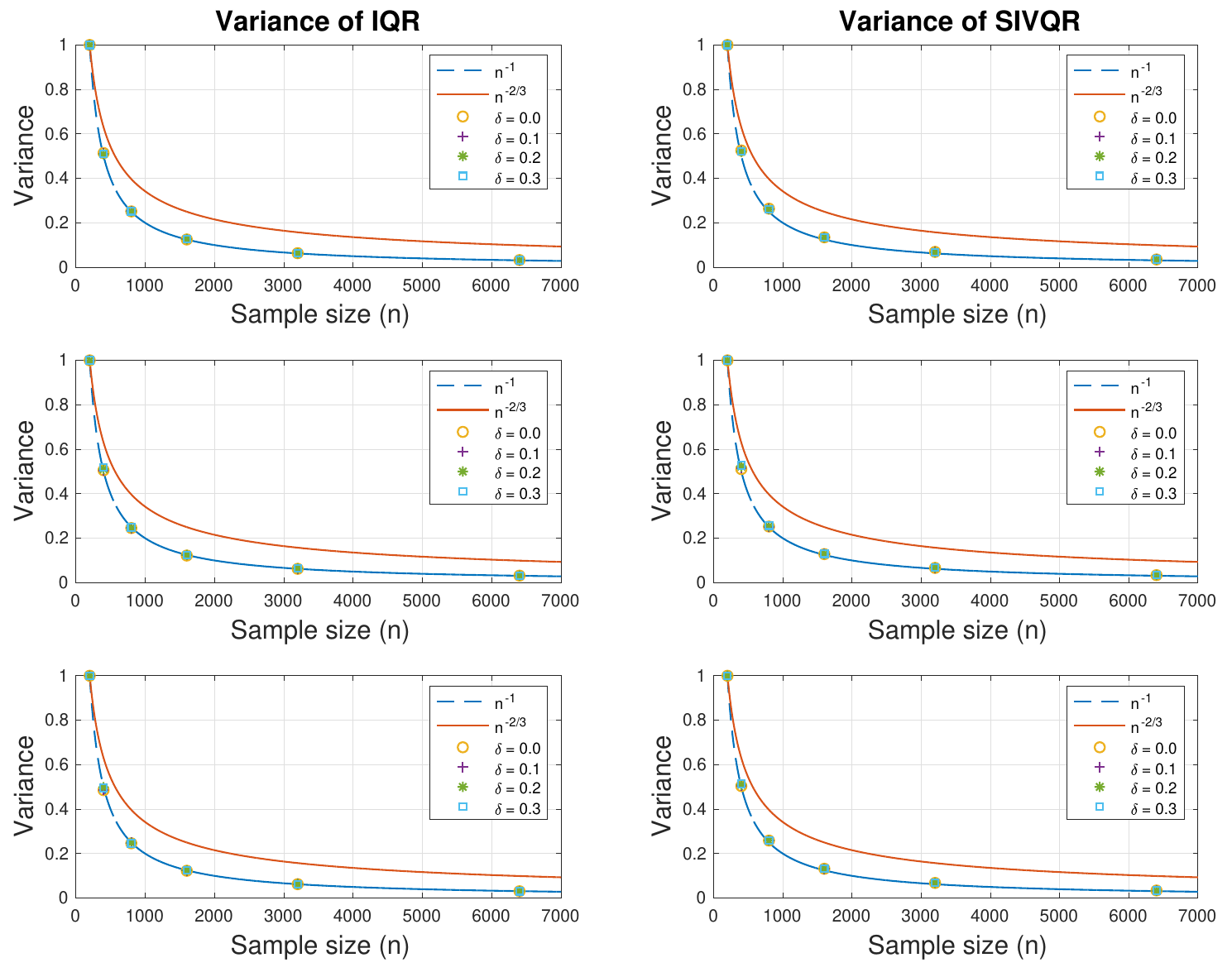}
    \caption{Variance Decay Rates of the IQR and SIVQR Estimators for $\beta_0(\tau)$}
    \label{fig2}
\end{figure}

Figure \ref{fig2} shows the variance of the IQR estimator in the left column and the variance of the SIVQR estimator in the right column. Each row corresponds to the results for the 0.25, 0.5, and 0.75 quantiles, respectively. We normalize the variance relative to its value at n = 200 for each $\tau$, n and $\delta$. We can confirm that, whether the model is correctly specified or misspecified, the variance of the estimator decreases inversely proportional to the sample size.

\section{Conclusion}

In this note, we investigated the convergence rate of the one-step and two-step GMM estimators with nonsmooth moment functions, considering potential misspecification of the moment condition model. Most results are consistent with theory. For directionally differentiable (e.g., check function) moment functions, the variance of GMM estimators decreases at the standard $n^{-1}$ rate regardless of misspecification (Hall and Inoue, 2003; Kang and Lee, 2024). In contrast, the variance of the GMM estimator with non-directionally differentiable (e.g., indicator function) moment functions decreases at the $n^{-2/3}$ rate under misspecification, confirming the theoretical findings of Hong and Li (2023).

However, there is one exception: our simulation results indicate that the variance of the one-step GMM estimator with the identity weight matrix decreases at the standard $n^{-1}$ rate, even under severe misspecification. The cause of this discrepancy between the theory and the finite-sample simulation results warrants further investigation.

\newpage
\section*{References}

\begin{description}
	\item[] Chen, Le?Yu, and Sokbae Lee. "Exact computation of GMM estimators for instrumental variable quantile regression models." \textit{Journal of Applied Econometrics} 33, no. 4 (2018): 553-567.
	\item[] Chernozhukov, Victor, and Christian Hansen. "An IV model of quantile treatment effects." \textit{Econometrica} 73, no. 1 (2005): 245-261.
	\item[] Chernozhukov, Victor, and Christian Hansen. "Instrumental quantile regression inference for structural and treatment effect models." \textit{Journal of Econometrics} 132, no. 2 (2006): 491-525.
	\item[] Chernozhukov, Victor, and Christian Hansen. "Instrumental variable quantile regression: A robust inference approach." \textit{Journal of Econometrics} 142, no. 1 (2008): 379-398.
	\item[] Hong, Han, and Jessie Li. "Rate-Adaptive Bootstrap for Possibly Misspecified GMM." \textit{Econometric Theory} (2023): 1-51.
	\item[] Hall, Alastair R., and Atsushi Inoue. "The large sample behaviour of the generalized method of moments estimator in misspecified models." \textit{Journal of Econometrics} 114, no. 2 (2003): 361-394.
	\item[] Hansen, Lars Peter. "Large sample properties of generalized method of moments estimators." \textit{Econometrica} (1982): 1029-1054.
	\item[] Hidalgo, Javier, Jungyoon Lee, and Myung Hwan Seo. "Robust inference for threshold regression models." \textit{Journal of Econometrics} 210, no. 2 (2019): 291-309.
	\item[] Kang, Byunghoon, and Seojeong Lee. "Robust Asymptotic and Bootstrap Inference for Nonsmooth GMM." Available at SSRN 4836402 (2024).
	\item[] Kaplan, David M., and Yixiao Sun. "Smoothed estimating equations for instrumental variables quantile regression." \textit{Econometric Theory} 33, no. 1 (2017): 105-157.
	\item[] Koo, Bonsoo, and Myung Hwan Seo. "Structural-break models under mis-specification: Implications for forecasting." \textit{Journal of Econometrics} 188, no. 1 (2015): 166-181.
	\item[] Machado, Jos\'{e} AF, and JMC Santos Silva. "Quantiles via moments." \textit{Journal of Econometrics} 213, no. 1 (2019): 145-173.
	\item[] Newey, Whitney K., and Daniel McFadden. "Large sample estimation and hypothesis testing." \textit{Handbook of Econometrics} 4 (1994): 2111-2245.
\end{description}

\end{document}